\documentclass[aps,prb,twocolumn,groupedaddress,showpacs,intlimits,amsmath,amssymb,floatfix,citeautoscript]{revtex4-1}

\usepackage{bm}
\usepackage[final]{graphicx}
\usepackage[hang,center]{subfigure}

\newcommand{\ueff}{$U_{\textrm{eff}}$ }

\begin{document}

\title{Competition between antiferromagnetic and charge density wave order in the half filled Hubbard-Holstein model}

\author{E. A. Nowadnick$^{1,2 *}$}
\author{S. Johnston$^{2,3,4 *}$}
\author{B. Moritz$^{2,5,6}$}
\author{R. T. Scalettar$^{7}$}
\author{T. P. Devereaux$^{2}$}
\affiliation{$^1$Department of Physics, Stanford University, Stanford, CA 94305, USA.}
\affiliation{$^2$Stanford Institute for Materials and Energy Sciences, SLAC National Accelerator Laboratory and 
Stanford University, Stanford, CA 94305, USA}
\affiliation{$^3$ Institute for Theoretical Solid State Physics, IFW Dresden, Helmholtzstrasse 20, 01069 Dresden, Germany}
\affiliation{$^4$ Department of Physics and Astronomy, University of Waterloo, Waterloo, ON N2L 3G1, Canada}
\affiliation{$^5$Department of Physics and Astrophysics, University of North Dakota, Grand Forks, ND 58202, USA}
\affiliation{$^6$Department of Physics, Northern Illinois University, DeKalb, IL 60115, USA}
\affiliation{$^7$ Department of Physics, University of California-- Davis, Davis, CA 95616, USA}

\date{\today}

\begin{abstract}
We present a determinant quantum Monte Carlo study of the competition between instantaneous on-site Coulomb repulsion and retarded phonon-mediated attraction between electrons, as described by the two dimensional Hubbard-Holstein model.  At half filling, we find a strong competition between antiferromagnetism (AFM) and charge density wave (CDW) order.  We demonstrate that  a simple picture
of AFM-CDW competition that incorporates the phonon mediated attraction into an effective-\emph{U} Hubbard model  requires significant refinement.  Specifically, retardation
effects slow the onset of charge order, so that CDW order remains absent even when the effective $U$ is negative.
This delay opens a window where neither AFM nor CDW order
is well established, and where there are signatures of a possible
metallic phase.
\end{abstract}

\pacs{71.10.Fd, 71.30.+h, 71.38.-k, 71.45.Lr, 74.72.-h} \maketitle

The electron-phonon (el-ph) interaction is responsible for many phenomena in condensed matter physics, including charge density waves (CDWs) and conventional superconductivity. While the el-ph interaction is well understood in metals, the role of phonons in strongly correlated systems is less clear, in part because the interplay of strong electron-electron (el-el) and el-ph interactions can lead to competing ordered phases. 
Despite its difficulty, this is an important problem to solve because multiple experimental probes have detected signatures of significant lattice effects in strongly correlated materials.  For example, in the cuprate high-temperature superconductors, angle-resolved photoemission spectroscopy (ARPES) has observed ``kinks" in the band dispersion, which have been attributed to the el-ph interaction,~\cite{lanzara} as well as small polaron formation in undoped Ca$_{2-x}$Na$_x$CuOCl$_2$.~\cite{shen,shen2}
Additional evidence for a significant el-ph interaction include strong quasiparticle renormalizations  detected by STM,~\cite{lee} and studies which have qualitatively reproduced optical conductivity peaks by including phonons.~\cite{mishchenko, deFilippis}  Besides the cuprates, other materials with both strong el-el and el-ph interactions include the manganites~\cite{millis} and fullerenes.~\cite{durand}

On general grounds, two effects are expected when el-ph interactions are included in a system with strong el-el repulsion.  The first is that the two interactions renormalize each other. The phonons mediate a retarded attractive el-el interaction, thus reducing the effective Coloumb repulsion, while the el-el repulsion suppresses charge fluctuations, and hence the el-ph interaction, which couples to them.  The second effect is a reduction in the quasiparticle weight due to additional scattering processes, which at large el-ph couplings can lead to a polaron crossover.  

A natural model for studying the interplay of the el-el and el-ph interactions is the Hubbard-Holstein (HH) model, which has been studied using various numerical approaches producing sometimes contradicting results.  Within dynamical mean field theory (DMFT), the suppression of the el-ph interaction  depends on the underlying phase, and antiferromagnetic (AFM)-DMFT has found  a moderate increase in the critical el-ph coupling for small polaron formation.~\cite{sangiovanni, sangiovanni2} In contrast,  diagrammatic quantum Monte Carlo work on the t-J-Holstein model found a reduction in the critical el-ph coupling needed for small polaron crossover.~\cite{andrey}  Dynamical cluster approximation (DCA) studies investigated the effect of phonons on the superconducting $T_c$, and found that  phonons suppress $T_c$ at small doping levels,~\cite{macridin} however, including longer range hopping terms in the presence of phonons enhanced $T_c$.~\cite{khatami} 


In addition to renormalization effects arising from the interplay of the el-el and el-ph interactions, competition between ordered phases can occur. On a two dimensional square lattice, at half filling the Hubbard and Holstein models have instabilities towards $(\pi/a,\pi/a)$ AFM  and CDW orders, respectively;  these phases compete in the HH model.  Due to the many-body nature of the problem, compounded by the many degrees of freedom in the HH model, in general there is no exact solution. In one dimension, the HH  phase diagram has been established via several numerical approaches, with an intermediate metallic state between the AFM and CDW phases.~\cite{takada,clay,hardikar,fehske,helen} The size of the metallic region grows with increasing phonon frequency.~\cite{clay, hardikar, fehske} A similar competition between AFM and CDW orders and phase diagram have been mapped out in infinite dimensions with DMFT.~\cite{koller,koller2,werner, bauer, bauer2} 
The AFM-CDW competition in two dimensions also has been studied with perturbative~\cite{kumar, berger} as well as strong coupling~\cite{hotta} techniques.

In this work we present a determinant quantum Monte Carlo (DQMC) study of the two dimensional single-band HH model at half filling.   DQMC is a numerically exact method that treats el-el and el-ph interactions on an equal footing and non-perturbatively. A nonzero el-ph coupling introduces the fermion sign problem at half filling.~\cite{steve_thesis}  Nevertheless, simulations for all
 parameter ranges presented in this work can be done down to
 $T=W/40$, where $W$ is the non-interacting bandwidth.
Significantly lower temperatures can be reached in some regimes.   For details of the DQMC method, please refer to Refs.~\onlinecite{BSS,White,scalettar_holstein}.

The Hamiltonian for the single-band HH model is $H = H_{kin}+H_{lat}+H_{int}$, where
\begin{eqnarray}
H_{kin}&=& -t \sum_{<ij>\sigma} c_{i\sigma}^\dagger c_{j\sigma}^{\phantom{\dagger}} -\mu \sum_{i\sigma}\hat{n}_{i\sigma} \\ \nonumber
H_{lat}&=& \sum_i \Big ( \frac{M\Omega^2}{2}\hat{X}_i^2+\frac{1}{2M}\hat{P}_i^2 \Big ) \\
H_{int}&=&U\sum_i \Big (\hat{n}_{i\uparrow}-\frac{1}{2} \Big )\Big  (\hat{n}_{i\downarrow}-\frac{1}{2} \Big ) -g\sum_{i\sigma}\hat{n}_{i\sigma}\hat{X}_i. \nonumber
\end{eqnarray}
Here $<$...$>$ denotes a sum over nearest neighbors, $c_{i\sigma}^{{\dagger}}$ creates an electron with spin $\sigma$ at site $i$, $\hat{n}_{i\sigma}=c_{i\sigma}^{{\dagger}} c_{i\sigma}^{\phantom{\dagger}}$, $t$ is the nearest neighbor hopping, $\Omega$ is the phonon frequency, $U$ is the el-el interaction strength, $g$ is the el-ph interaction strength, and $\mu$ is the chemical potential, which is adjusted to maintain half filling. The dimensionless electron-phonon coupling constant is defined as $\lambda=g^2/M\Omega^2 W$. Throughout we take $t=1$, $M=1$,  and $a=1$ as our units of energy, mass, and length, respectively. 

 We first study the spin and charge susceptibilities $\chi_s$ and $\chi_c$, which are given by 
\begin{equation}
\chi_{s,c}({\bf q})=\frac{1}{N} \int_0^\beta d\tau <T_\tau \hat{O}^{\phantom{\dagger}}_{s,c}({\bf q},\tau) \hat{O}^\dagger_{s,c}({\bf q},0)>
\end{equation}
where $\hat{O}_s({\bf q})=\sum_i e^{i{\bf q}\cdot {\bf R}_i}(\hat{n}_{i\uparrow}-\hat{n}_{i\downarrow})$, and $\hat{O}_c({\bf q})=\sum_{i} e^{i{\bf q}\cdot {\bf R}_i}(\hat{n}_{i\uparrow}+\hat{n}_{i\downarrow})$.

\begin{figure}
\includegraphics[width=0.5\textwidth]{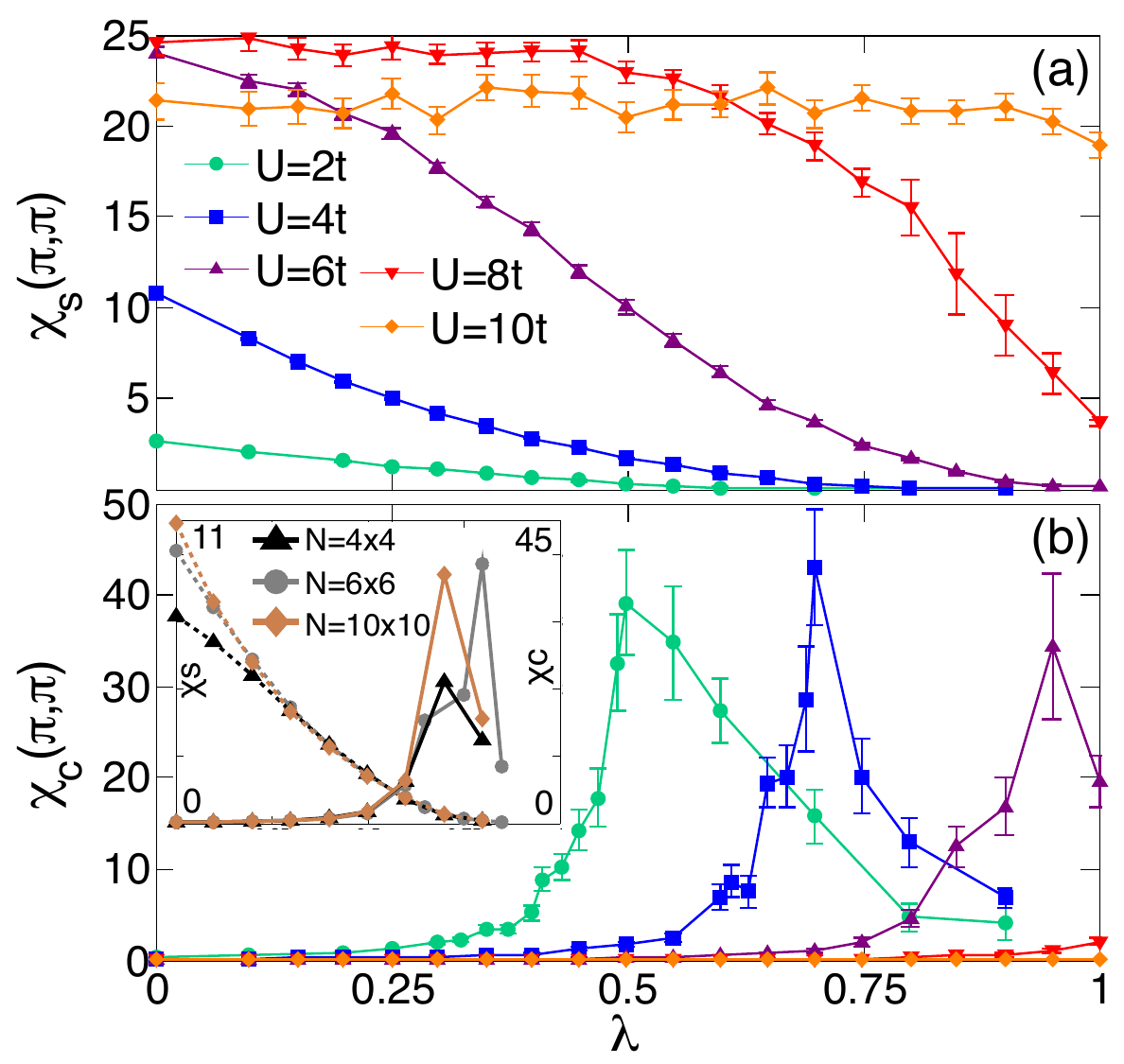}
\caption{\label{fig:susc} (a)  $\chi_{s}(\pi,\pi)$ and (b) $\chi_{c}(\pi,\pi)$ for several $U$ on an $N=8\times 8$ lattice.  Inset of  (b) shows  $\chi_s$ (dashed lines) and $\chi_c$ (solid lines) at $U=4t$ for several lattice sizes. The error bars in the inset are suppressed for clarity.  The remaining simulation parameters are: $\beta=4/t$, $\Delta \tau=0.1/t$, $\Omega=t$. }
\end{figure}

The susceptibilities  at wavevector ${\bf q}=(\pi, \pi)$ are shown in Fig.~\ref{fig:susc} for several values of $U$. With increasing $\lambda$, $\chi_s$ decreases, signaling that the el-ph interaction reduces the strength of the effective el-el repulsion.  This decrease in $\chi_{s}$ occurs immediately with the inclusion of nonzero $\lambda$ for low to intermediate $U$, while for large $U$, suppression of $\chi_{s}$ does not occur until the el-ph coupling is fairly strong ($\lambda$=0.5 for $U$=8$t$, and $\lambda>1$ for $U$=10$t$). As $\chi_{s}$ shrinks, $\chi_{c}$ increases, indicating a clear competition between the spin and charge orders.  For all values of $U$ considered here, $\chi_c$ is negligible up to a $U$-dependent critical  $\lambda$, at which point it grows rapidly. However, for strong el-el interactions ($U=8t$, $10t$), $\chi_{c}$ is still relatively small even at $\lambda=1$, due to the strong tendency toward AFM still present. Interestingly, rather than continuously growing with $\lambda$, the CDW susceptibility peaks and then decreases,  for $U=2t$-$6t$.  We attribute this behavior to the finite CDW transition temperature in the HH model, which will be discussed in more detail below. The inset in Fig.~\ref{fig:susc}(b) shows $\chi_s$ and $\chi_c$ for $U=4t$ for several lattice sizes, demonstrating that the lattice size has little effect on our conclusions.  

\begin{figure}
\includegraphics[width=0.5\textwidth]{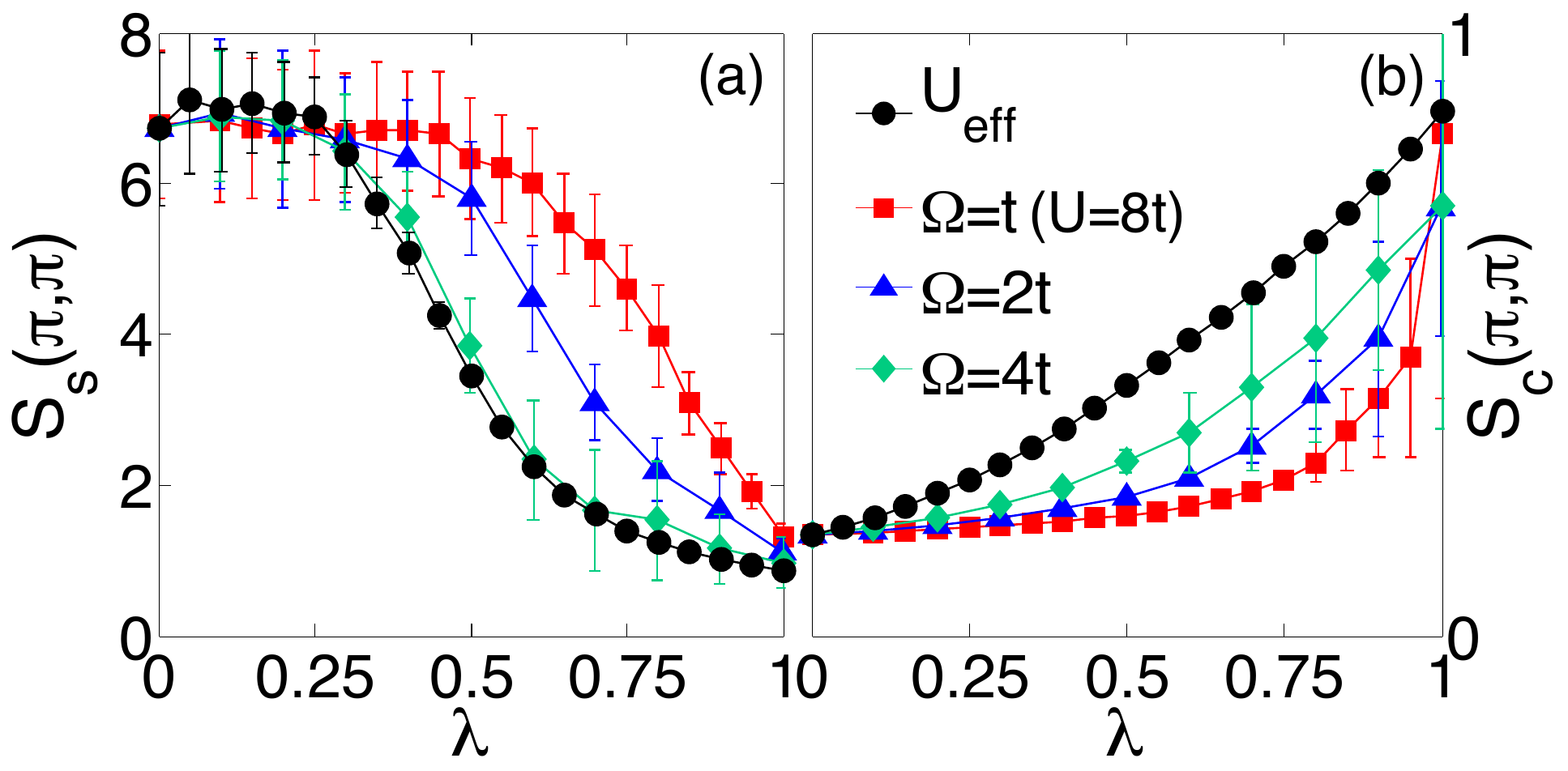}
\caption{\label{fig:ueff} Structure factors (a) $S_{s}(\pi,\pi)$ and (b) $S_{c}(\pi,\pi)$  for a \ueff Hubbard model (black) and the $U=8t$ HH model with several phonon frequencies. The remaining simulation parameters are: $N=8\times 8$, $\beta=4/t$, $\Delta \tau=0.1/t$. }
\end{figure}

Since one of the effects of el-ph coupling is to reduce the effective strength of $U$, we investigate how well a \ueff    Hubbard model can describe the physics of the HH model. Integrating out the phonon field in the HH model yields a dynamic el-el interaction:
  \begin{equation}
 U_{\textrm{eff}}(\omega)=U-\frac{g^2}{M(\Omega^2-\omega^2)}.
 \end{equation}
In the anti-adiabatic limit ($\Omega$$\rightarrow$$\infty$) this interaction becomes instantaneous, and reduces to the  form $U_{\textrm{eff}}=U-\lambda W$. A frequency-independent \ueff Hubbard model is often used to describe the HH model, even at finite $\Omega$. For example, DMFT studies have found that such an approach captures the low-energy physics of the HH model.~\cite{sangiovanni, bauer}

Fig.~\ref{fig:ueff} compares the spin and charge structure factors $S_{s,c}({\bf q})=<$$\hat{O}^{\phantom{\dagger}}_{s,c}({\bf q})\hat{O}^\dagger_{s,c}({\bf q})$$>$ at  ${\bf q}=(\pi,\pi)$ of a frequency-independent \ueff Hubbard model and the $U=8t$ HH model at several phonon frequencies.  Up to $\lambda \approx 0.25$, $S_{s}(\pi,\pi)$ in the \ueff and HH models agrees for all $\Omega$ considered.  Beyond this point, $S_{s}(\pi,\pi)$ is suppressed more slowly in the HH model than in the \ueff model, due to the retarded nature of the el-ph interaction captured in HH. As $\Omega$ increases,  the HH result comes closer to the \ueff result, until by $\Omega=4t$, the two models agree within the error bars. We also considered other values of $U$ (not shown), and found that for a given $\Omega$ the difference between the HH and \ueff results grows as  $U$ increases.  In contrast to $S_{s}(\pi,\pi)$, $S_{c}(\pi,\pi)$ calculated in the \ueff and HH models does not agree for any  $\lambda$.  Rather, $S_{c}(\pi,\pi)$  immediately rises in the \ueff model, while in the HH model it remains small until  $\lambda \approx 0.75$, and then rises sharply.  As the phonon frequency increases, the HH and \ueff results get closer, although they are still inconsistent at $\Omega=4t$.  This result is generic; while the \ueff Hubbard model has a CDW phase for any \ueff that is negative, $\chi_c$ remains suppressed well beyond the $\lambda$ at which \ueff =0, as is clear  in Fig.~\ref{fig:susc}(b) where \ueff=0 at $\lambda$= 0.25, 0.5, and 0.75 for $U$=$2t$, $4t$, and $6t$, respectively.

An additional difference between the HH and \ueff models is the CDW transition temperature. In the HH model, while $T_{AFM}=0$ in two dimensions due to the Mermin-Wagner theorem, $T_{CDW}$ is finite because the  order parameter has two states.  DQMC finite size scaling studies~\cite{noack,vekic} of the Holstein model found that $t\beta_{CDW}=8-11$ for $\Omega=t$ and $\lambda=0.25$.  While we did not perform a scaling analysis for the HH model, we expect $T_{CDW}$ to be in the same temperature regime, because while
the inclusion of $U$ in the HH model localizes carriers (which would lower $T_{CDW}$), it also pushes the CDW transition to a larger $\lambda$ (which would increase $T_{CDW}$).  In contrast,  $T_{CDW}=0$ in the attractive-$U$ Hubbard model. The sharply peaked nature of $\chi_c$ in Fig.~\ref{fig:susc}(b), differing from the slow evolution of $\chi_s$, may be due to the proximity of the temperature $t\beta=4$ to $T_{CDW}$.

We now turn to the spectral properties of the HH model. To avoid analytic continuation, we focus on the spectral weight near the Fermi level, which is obtained  from the imaginary time propagator via the relation~\cite{trivedi}
\begin{equation}
\beta C({\bf k},\tau=\beta/2)=\frac{\beta}{2}\int d\omega A({\bf k},\omega)g(\omega,\beta)
\end{equation}
where $C$ and $A$ are the propagator and spectral function, respectively, and $g(\omega,\beta)=\omega/\sinh(\beta\omega/2)$ for bosons and $=1/\cosh(\beta\omega/2)$ for fermions. At low temperature, $g(\omega,\beta)$ is sharply peaked about $\omega=0$, so that $A({\bf k},\omega=0)$ dominates the integral. We consider the local propagator $C({\bf r}=0)=\sum_{\bf k}C({\bf k})$, which is related to the low energy projected density of states via $N(0) \approx \beta C({\bf r}=0,\tau=\beta/2)/\pi$. 

\begin{figure}
\includegraphics[width=0.5\textwidth]{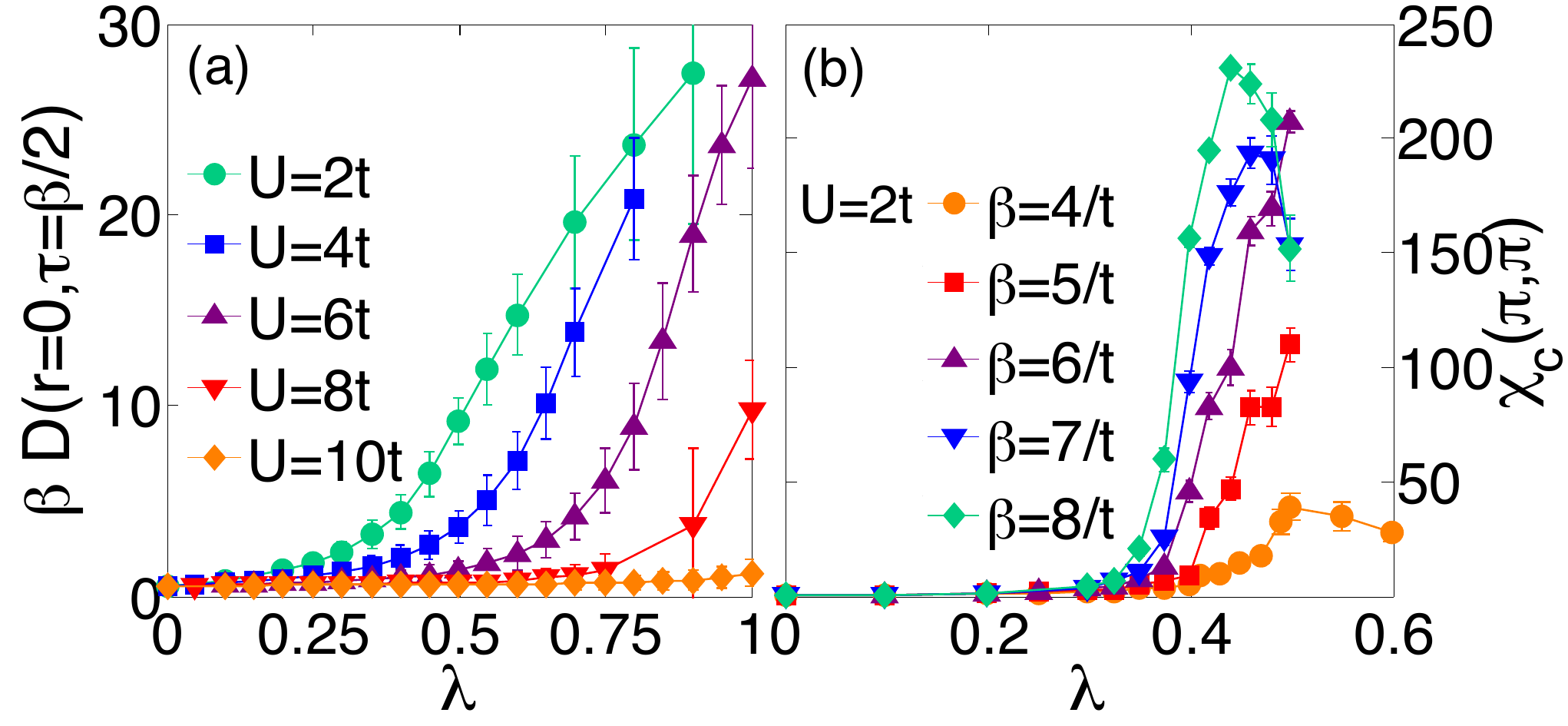}
\caption{\label{fig:phonon} (a) Local  phonon propagator  $\beta D({\bf r}=0,\tau=\beta/2)$ for several values of $U$ at $\beta=4/t$. (b) $\chi_c(\pi,\pi)$ for $U=2t$ and several values of $\beta.$ The remaining simulation parameters are: $N=8\times 8$,  $\Delta \tau=0.1/t$, $\Omega=t$. }
\end{figure}

The phonon propagator,  defined as $D_{ij}(\tau)$= $<$$T_\tau \hat{X}_i(\tau) \hat{X}_j(0)$$>$$-$$<$$X$$>^2$, contains information on phonon softening at the CDW transition. In the Holstein model, the phonon spectral function  is peaked at the bare phonon frequency $\pm \Omega$ in a system without el-ph coupling; el-ph interactions renormalize the phonon frequency and lead to spectral weight at other frequencies.  In particular, the appearance of spectral weight at $\omega=0$ indicates the development of a static CDW lattice distortion, which is revealed by $\beta D({\bf r}=0,\tau=\beta/2)$ (abbreviated as $\beta D_{\beta/2}$), as shown in Fig.~\ref{fig:phonon}.  For low el-ph coupling, $\beta D_{\beta/2}$ is negligible, since the system is far from the CDW state.  It then increases at the same $U$-dependent  $\lambda$ at which $\chi_{c}$ rapidly increases in Fig.~\ref{fig:susc}(b). This phonon softening indicates that the CDW formation may have a Peierls-like origin, in which case the Fermi surface could be restored  during the transition from an AFM to a CDW insulator. Fig.~\ref{fig:phonon}(b) shows $\chi_c$ at $U=2t$ for several lower temperatures.  With decreasing temperature, the rise in $\chi_c$ sharpens dramatically and also shifts to lower $\lambda$, appearing to asymptote towards a divergence in the susceptibility at low temperature around $\lambda=0.3$.  We also note that the peak and subsequent decay in the CDW susceptibility discussed earlier appears robustly as a function of temperature.

 \begin{figure*}
 \begin{center}
\includegraphics[width=0.7\textwidth]{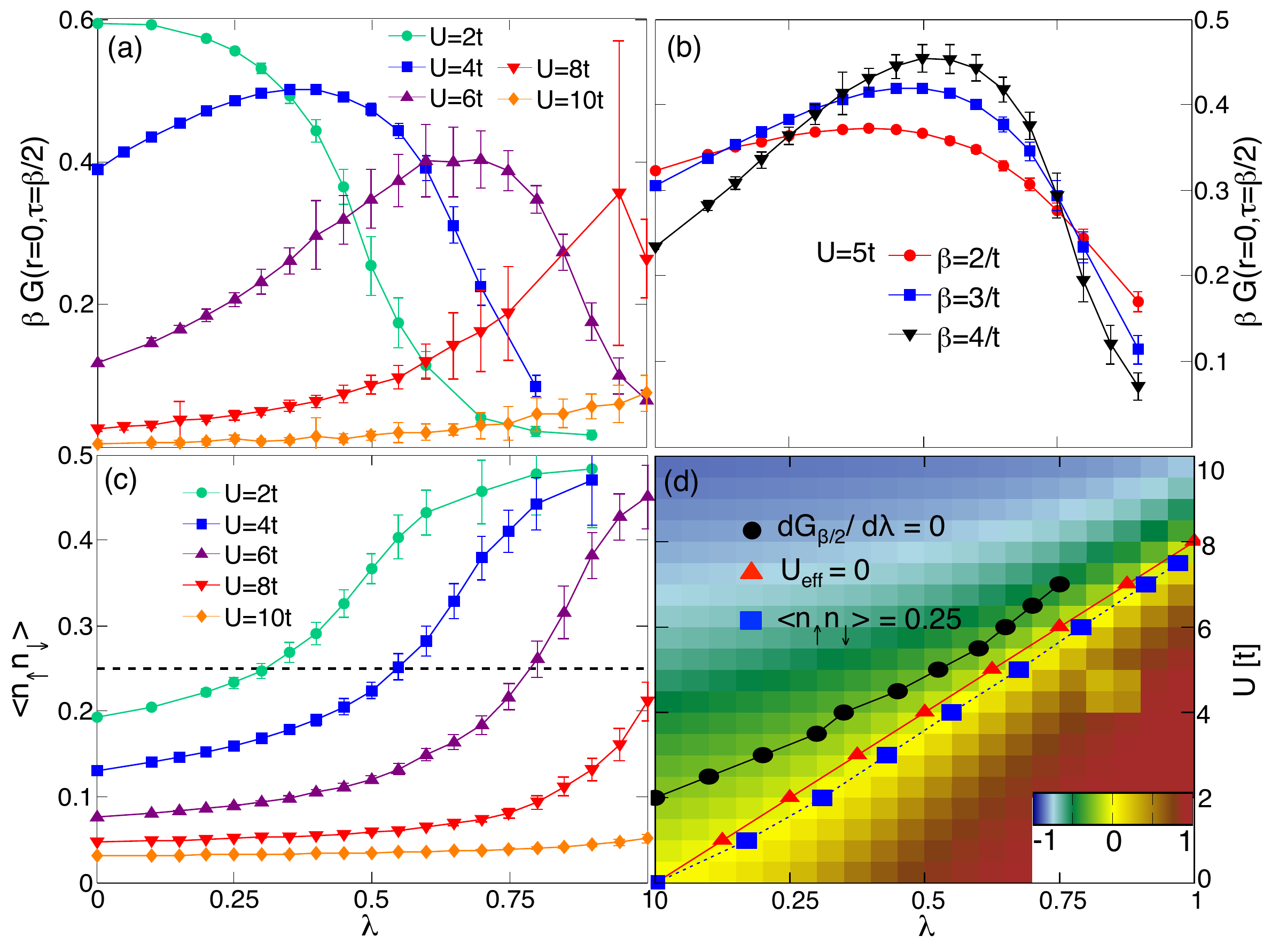}
\caption{\label{fig:pd}  Local electron Green's function $\beta G({\bf r}=0,\tau=\beta/2)$ for (a) several values of $U$ at $\beta=4/t$, and (b) at $U=5t$ for several $\beta$.  (c) Average double occupancy $<$$n_\uparrow n_\downarrow$$>$  at $\beta=4/t$. The dashed line indicates where $<$$n_\uparrow n_\downarrow$$>$=0.25. (d) $\Phi_c-\Phi_s$ at $\beta=4/t$. The remaining simulation parameters are: $N=8\times 8$,  $\Delta \tau=0.1/t$, $\Omega=t$.}
\end{center}
\end{figure*}

The electronic spectral weight near $\omega=0$ also offers insight into the AFM-CDW transition.  In this case, $\beta G({\bf r}=0, \tau=\beta/2)$ (abbreviated as $\beta G_{\beta/2}$) distinguishes between insulating and metallic systems, being 0 in the low temperature limit when a gap is present, and finite if a band disperses through the Fermi level. 
 Fig.~\ref{fig:pd}(a) shows $\beta G_{\beta/2}$ for several values of $U$.  For small $\lambda$, $\beta G_{\beta/2}$ decreases with increasing $U$, indicating the opening of the Mott gap.  As a function of increasing $\lambda$, $\beta G_{\beta/2}$ falls for $U$=$2t$ as the CDW gap develops. For $U=4t$ and $6t$, $\beta G_{\beta/2}$ initially grows  as the el-ph interaction reduces the effective el-el repulsion and the Mott gap closes, and then decreases quickly as the CDW gap opens. For all these $U$,  $\beta G_{\beta/2}$ begins to fall at the same $\lambda$ at which $\chi_c$ increases in Fig.~\ref{fig:susc}(b) and the phonon softens in Fig.~\ref{fig:phonon}(a). For $U=8t$ and $10t$ $\beta G_{\beta/2}$ grows slowly with $\lambda$ as the Mott gap narrows.
  
 What can the peak in $\beta G_{\beta/2}$ at intermediate $\lambda$ in Fig.~\ref{fig:pd}(a) tell us about the AFM-CDW transition?  In Fig.~\ref{fig:pd}(b), we plot $\beta G_{\beta/2}$ at $U=5t$ for several temperatures.  As the temperature is lowered, $\beta G_{\beta/2}$ decreases at small and large $\lambda$ as the Mott and CDW gaps open, respectively.  However, at intermediate el-ph coupling, $\beta G_{\beta/2}$ actually grows with decreasing temperature, behavior that could arise from an intervening metallic phase between the Mott and CDW insulating states.  This increase in spectral weight at intermediate $\lambda$ was observed robustly for several  lattice sizes and values of $U$. In addition, the possible implication of a Fermi surface in the intermediate state, from the phonon softening in Fig.~\ref{fig:phonon}(a), further supports the idea of an intermediate metallic phase.
 
To further explore signatures of this possible metallic state, we plot the  average double occupancy $<$$n_\uparrow n_\downarrow$$>$ in Fig.~\ref{fig:pd}(c).  The double occupancy distinguishes between $(\pi,\pi)$ AFM and CDW insulators, where it is 0 and 0.5, respectively.  In a metallic state (or an AFM-CDW coexistence state), $<$$n_\uparrow n_\downarrow$$>=0.25$.  We find that $<$$n_\uparrow n_\downarrow$$>$ increases smoothly with energy through 0.25, which is consistent with an intermediate metallic state, rather than a direct AFM-CDW transition at a critical $\lambda$, where a sharp jump would be expected.  
While the transition from low to high double occupancy may sharpen as temperature is lowered, we note that in the range $t\beta=2-5$, we found much less temperature dependence in $<$$n_\uparrow n_\downarrow$$>$ than in other quantities considered in this paper.

A finite temperature $U-\lambda$ phase diagram for $\beta=4/t$, depicting the difference of the charge and spin order parameters, $\Phi_c-\Phi_s$, is shown in Fig.~\ref{fig:pd}(d). Here, the order parameters are defined as $\Phi_s=\sum_i(\hat{n}_{i\uparrow}-\hat{n}_{i\downarrow})^2/N$ and $\Phi_c=\sum_{i\sigma}(\hat{n}_{i\sigma}-1)^2/N$. Lines denoting \ueff=0, $<$$n_\uparrow n_\downarrow$$>=0.25$, and the peak in $\beta G_{\beta/2}$ are also included.  The dominance of AFM and CDW orders at large $U$ and $\lambda$, respectively, is apparent.  However, a sizable transition region, where $\Phi_c-\Phi_s \approx 0$ is clearly visible. 
The lines \ueff=0 and  $<$$n_\uparrow n_\downarrow$$>=0.25$ lie in the center of the transition region, while the peak in $\beta G_{\beta/2}$ is toward the side dominated by spin order. The coincidence of multiple quantities consistent with a metallic state in the intermediate region of the phase diagram corroborate the case for the existence of such a phase. 


To summarize, in this work we demonstrated a strong competition between AFM and CDW phases in the two dimensional single band HH model, and found evidence for a possible intermediate metallic regime existing between the ordered phases.  We investigated how well an effective-$U$ Hubbard model can describe the physics of the HH model, and found that while in some regimes the two models give comparable results, in general the retarded nature of the el-ph interaction leads to significant differences.  The $U$-$\lambda$  phase diagram determined in our study is qualitatively similar to that found by low temperature numerical approaches, with the \ueff=0 line dividing the regions of dominant spin and charge order parameters.  We found evidence for an intermediate metallic phase in two dimensions, similar to previous 1-$d$ results.~\cite{takada,clay,hardikar,fehske}. The size of the intermediate metallic region shrinks as the interaction strengths grow, which is consistent with Refs.~\onlinecite{clay,hardikar,fehske} where the metallic phase is found to terminate at strong couplings.  These findings contrast with the infinite dimensional DMFT results of Refs.~\onlinecite{bauer, bauer2}, where a direct order to order transition was found.  Potential explorations for future work include studying the effect of phonon frequency on the intermediate metallic state, and understanding more precisely where the metal-insulator transitions lie.

$Acknowledgements:$ * E.A.N. and S.J. contributed equally to this work. We thank N. Nagaosa and A. S. Mishchenko for useful discussions. We acknowledge support 
from  the U. S. Department of Energy, 
Office of Basic Energy Sciences, Materials Science and 
Engineering Division under Contract Numbers DE-AC02-76SF00515 and DE-FC0206ER25793. E.A.N. acknowledges support from EAPSI and S. J. acknowledges support from SHARCNET and NSERC.

\newpage

\bibliographystyle{apsrev}
\bibliography{nowadnick}

\begin{thebibliography}{33}
\expandafter\ifx\csname natexlab\endcsname\relax\def\natexlab#1{#1}\fi
\expandafter\ifx\csname bibnamefont\endcsname\relax
  \def\bibnamefont#1{#1}\fi
\expandafter\ifx\csname bibfnamefont\endcsname\relax
  \def\bibfnamefont#1{#1}\fi
\expandafter\ifx\csname citenamefont\endcsname\relax
  \def\citenamefont#1{#1}\fi
\expandafter\ifx\csname url\endcsname\relax
  \def\url#1{\texttt{#1}}\fi
\expandafter\ifx\csname urlprefix\endcsname\relax\def\urlprefix{URL }\fi
\providecommand{\bibinfo}[2]{#2}
\providecommand{\eprint}[2][]{\url{#2}}

\bibitem[{\citenamefont{{Lanzara} et~al.}(2001)\citenamefont{{Lanzara},
  {Bogdanov}, {Zhou}, {Kellar}, {Feng}, {Lu}, {Yoshida}, {Eisaki}, {Fujimori},
  {Kishio} et~al.}}]{lanzara}
\bibinfo{author}{\bibfnamefont{A.}~\bibnamefont{{Lanzara}}},
  \bibinfo{author}{\bibfnamefont{P.~V.} \bibnamefont{{Bogdanov}}},
  \bibinfo{author}{\bibfnamefont{X.~J.} \bibnamefont{{Zhou}}},
  \bibinfo{author}{\bibfnamefont{S.~A.} \bibnamefont{{Kellar}}},
  \bibinfo{author}{\bibfnamefont{D.~L.} \bibnamefont{{Feng}}},
  \bibinfo{author}{\bibfnamefont{E.~D.} \bibnamefont{{Lu}}},
  \bibinfo{author}{\bibfnamefont{T.}~\bibnamefont{{Yoshida}}},
  \bibinfo{author}{\bibfnamefont{H.}~\bibnamefont{{Eisaki}}},
  \bibinfo{author}{\bibfnamefont{A.}~\bibnamefont{{Fujimori}}},
  \bibinfo{author}{\bibfnamefont{K.}~\bibnamefont{{Kishio}}},
  \bibnamefont{et~al.}, \bibinfo{journal}{\nat} \textbf{\bibinfo{volume}{412}},
  \bibinfo{pages}{510} (\bibinfo{year}{2001}).

\bibitem[{\citenamefont{Shen et~al.}(2004)\citenamefont{Shen, Ronning, Lu, Lee,
  Ingle, Meevasana, Baumberger, Damascelli, Armitage, Miller et~al.}}]{shen}
\bibinfo{author}{\bibfnamefont{K.~M.} \bibnamefont{Shen}},
  \bibinfo{author}{\bibfnamefont{F.}~\bibnamefont{Ronning}},
  \bibinfo{author}{\bibfnamefont{D.~H.} \bibnamefont{Lu}},
  \bibinfo{author}{\bibfnamefont{W.~S.} \bibnamefont{Lee}},
  \bibinfo{author}{\bibfnamefont{N.~J.~C.} \bibnamefont{Ingle}},
  \bibinfo{author}{\bibfnamefont{W.}~\bibnamefont{Meevasana}},
  \bibinfo{author}{\bibfnamefont{F.}~\bibnamefont{Baumberger}},
  \bibinfo{author}{\bibfnamefont{A.}~\bibnamefont{Damascelli}},
  \bibinfo{author}{\bibfnamefont{N.~P.} \bibnamefont{Armitage}},
  \bibinfo{author}{\bibfnamefont{L.~L.} \bibnamefont{Miller}},
  \bibnamefont{et~al.}, \bibinfo{journal}{Phys. Rev. Lett.}
  \textbf{\bibinfo{volume}{93}}, \bibinfo{pages}{267002}
  (\bibinfo{year}{2004}).

\bibitem[{\citenamefont{Shen et~al.}(2007)\citenamefont{Shen, Ronning,
  Meevasana, Lu, Ingle, Baumberger, Lee, Miller, Kohsaka, Azuma
  et~al.}}]{shen2}
\bibinfo{author}{\bibfnamefont{K.~M.} \bibnamefont{Shen}},
  \bibinfo{author}{\bibfnamefont{F.}~\bibnamefont{Ronning}},
  \bibinfo{author}{\bibfnamefont{W.}~\bibnamefont{Meevasana}},
  \bibinfo{author}{\bibfnamefont{D.~H.} \bibnamefont{Lu}},
  \bibinfo{author}{\bibfnamefont{N.~J.~C.} \bibnamefont{Ingle}},
  \bibinfo{author}{\bibfnamefont{F.}~\bibnamefont{Baumberger}},
  \bibinfo{author}{\bibfnamefont{W.~S.} \bibnamefont{Lee}},
  \bibinfo{author}{\bibfnamefont{L.~L.} \bibnamefont{Miller}},
  \bibinfo{author}{\bibfnamefont{Y.}~\bibnamefont{Kohsaka}},
  \bibinfo{author}{\bibfnamefont{M.}~\bibnamefont{Azuma}},
  \bibnamefont{et~al.}, \bibinfo{journal}{Phys. Rev. B}
  \textbf{\bibinfo{volume}{75}}, \bibinfo{pages}{075115}
  (\bibinfo{year}{2007}).

\bibitem[{\citenamefont{{Lee} et~al.}(2006)\citenamefont{{Lee}, {Fujita},
  {McElroy}, {Slezak}, {Wang}, {Aiura}, {Bando}, {Ishikado}, {Masui}, {Zhu}
  et~al.}}]{lee}
\bibinfo{author}{\bibfnamefont{J.}~\bibnamefont{{Lee}}},
  \bibinfo{author}{\bibfnamefont{K.}~\bibnamefont{{Fujita}}},
  \bibinfo{author}{\bibfnamefont{K.}~\bibnamefont{{McElroy}}},
  \bibinfo{author}{\bibfnamefont{J.~A.} \bibnamefont{{Slezak}}},
  \bibinfo{author}{\bibfnamefont{M.}~\bibnamefont{{Wang}}},
  \bibinfo{author}{\bibfnamefont{Y.}~\bibnamefont{{Aiura}}},
  \bibinfo{author}{\bibfnamefont{H.}~\bibnamefont{{Bando}}},
  \bibinfo{author}{\bibfnamefont{M.}~\bibnamefont{{Ishikado}}},
  \bibinfo{author}{\bibfnamefont{T.}~\bibnamefont{{Masui}}},
  \bibinfo{author}{\bibfnamefont{J.-X.} \bibnamefont{{Zhu}}},
  \bibnamefont{et~al.}, \bibinfo{journal}{\nat} \textbf{\bibinfo{volume}{442}},
  \bibinfo{pages}{546} (\bibinfo{year}{2006}).

\bibitem[{\citenamefont{Mishchenko et~al.}(2008)\citenamefont{Mishchenko,
  Nagaosa, Shen, De~Filippis, Cataudella, Devereaux, Bernhard, Kim, and
  Zaanen}}]{mishchenko}
\bibinfo{author}{\bibfnamefont{A.~S.} \bibnamefont{Mishchenko}},
  \bibinfo{author}{\bibfnamefont{N.}~\bibnamefont{Nagaosa}},
  \bibinfo{author}{\bibfnamefont{Z.-X.} \bibnamefont{Shen}},
  \bibinfo{author}{\bibfnamefont{G.}~\bibnamefont{De~Filippis}},
  \bibinfo{author}{\bibfnamefont{V.}~\bibnamefont{Cataudella}},
  \bibinfo{author}{\bibfnamefont{T.~P.} \bibnamefont{Devereaux}},
  \bibinfo{author}{\bibfnamefont{C.}~\bibnamefont{Bernhard}},
  \bibinfo{author}{\bibfnamefont{K.~W.} \bibnamefont{Kim}}, \bibnamefont{and}
  \bibinfo{author}{\bibfnamefont{J.}~\bibnamefont{Zaanen}},
  \bibinfo{journal}{Phys. Rev. Lett.} \textbf{\bibinfo{volume}{100}},
  \bibinfo{pages}{166401} (\bibinfo{year}{2008}).

\bibitem[{\citenamefont{De~Filippis et~al.}(2009)\citenamefont{De~Filippis,
  Cataudella, Mishchenko, Perroni, and Nagaosa}}]{deFilippis}
\bibinfo{author}{\bibfnamefont{G.}~\bibnamefont{De~Filippis}},
  \bibinfo{author}{\bibfnamefont{V.}~\bibnamefont{Cataudella}},
  \bibinfo{author}{\bibfnamefont{A.~S.} \bibnamefont{Mishchenko}},
  \bibinfo{author}{\bibfnamefont{C.~A.} \bibnamefont{Perroni}},
  \bibnamefont{and} \bibinfo{author}{\bibfnamefont{N.}~\bibnamefont{Nagaosa}},
  \bibinfo{journal}{Phys. Rev. B} \textbf{\bibinfo{volume}{80}},
  \bibinfo{pages}{195104} (\bibinfo{year}{2009}).

\bibitem[{\citenamefont{{Millis}}(1998)}]{millis}
\bibinfo{author}{\bibfnamefont{A.~J.} \bibnamefont{{Millis}}},
  \bibinfo{journal}{\nat} \textbf{\bibinfo{volume}{392}}, \bibinfo{pages}{147}
  (\bibinfo{year}{1998}).

\bibitem[{\citenamefont{{Durand} et~al.}(2003)\citenamefont{{Durand},
  {Darling}, {Dubitsky}, {Zaopo}, and {Rosseinsky}}}]{durand}
\bibinfo{author}{\bibfnamefont{P.}~\bibnamefont{{Durand}}},
  \bibinfo{author}{\bibfnamefont{G.~R.} \bibnamefont{{Darling}}},
  \bibinfo{author}{\bibfnamefont{Y.}~\bibnamefont{{Dubitsky}}},
  \bibinfo{author}{\bibfnamefont{A.}~\bibnamefont{{Zaopo}}}, \bibnamefont{and}
  \bibinfo{author}{\bibfnamefont{M.~J.} \bibnamefont{{Rosseinsky}}},
  \bibinfo{journal}{Nature Materials} \textbf{\bibinfo{volume}{2}},
  \bibinfo{pages}{605} (\bibinfo{year}{2003}).

\bibitem[{\citenamefont{Sangiovanni et~al.}(2005)\citenamefont{Sangiovanni,
  Capone, Castellani, and Grilli}}]{sangiovanni}
\bibinfo{author}{\bibfnamefont{G.}~\bibnamefont{Sangiovanni}},
  \bibinfo{author}{\bibfnamefont{M.}~\bibnamefont{Capone}},
  \bibinfo{author}{\bibfnamefont{C.}~\bibnamefont{Castellani}},
  \bibnamefont{and} \bibinfo{author}{\bibfnamefont{M.}~\bibnamefont{Grilli}},
  \bibinfo{journal}{Phys. Rev. Lett.} \textbf{\bibinfo{volume}{94}},
  \bibinfo{pages}{026401} (\bibinfo{year}{2005}).

\bibitem[{\citenamefont{Sangiovanni et~al.}(2006)\citenamefont{Sangiovanni,
  Gunnarsson, Koch, Castellani, and Capone}}]{sangiovanni2}
\bibinfo{author}{\bibfnamefont{G.}~\bibnamefont{Sangiovanni}},
  \bibinfo{author}{\bibfnamefont{O.}~\bibnamefont{Gunnarsson}},
  \bibinfo{author}{\bibfnamefont{E.}~\bibnamefont{Koch}},
  \bibinfo{author}{\bibfnamefont{C.}~\bibnamefont{Castellani}},
  \bibnamefont{and} \bibinfo{author}{\bibfnamefont{M.}~\bibnamefont{Capone}},
  \bibinfo{journal}{Phys. Rev. Lett.} \textbf{\bibinfo{volume}{97}},
  \bibinfo{pages}{046404} (\bibinfo{year}{2006}).

\bibitem[{\citenamefont{Mishchenko and Nagaosa}(2004)}]{andrey}
\bibinfo{author}{\bibfnamefont{A.~S.} \bibnamefont{Mishchenko}}
  \bibnamefont{and} \bibinfo{author}{\bibfnamefont{N.}~\bibnamefont{Nagaosa}},
  \bibinfo{journal}{Phys. Rev. Lett.} \textbf{\bibinfo{volume}{93}},
  \bibinfo{pages}{036402} (\bibinfo{year}{2004}).

\bibitem[{\citenamefont{Macridin et~al.}(2006)\citenamefont{Macridin, Moritz,
  Jarrell, and Maier}}]{macridin}
\bibinfo{author}{\bibfnamefont{A.}~\bibnamefont{Macridin}},
  \bibinfo{author}{\bibfnamefont{B.}~\bibnamefont{Moritz}},
  \bibinfo{author}{\bibfnamefont{M.}~\bibnamefont{Jarrell}}, \bibnamefont{and}
  \bibinfo{author}{\bibfnamefont{T.}~\bibnamefont{Maier}},
  \bibinfo{journal}{Phys. Rev. Lett.} \textbf{\bibinfo{volume}{97}},
  \bibinfo{pages}{056402} (\bibinfo{year}{2006}).

\bibitem[{\citenamefont{Khatami et~al.}(2008)\citenamefont{Khatami, Macridin,
  and Jarrell}}]{khatami}
\bibinfo{author}{\bibfnamefont{E.}~\bibnamefont{Khatami}},
  \bibinfo{author}{\bibfnamefont{A.}~\bibnamefont{Macridin}}, \bibnamefont{and}
  \bibinfo{author}{\bibfnamefont{M.}~\bibnamefont{Jarrell}},
  \bibinfo{journal}{Phys. Rev. B} \textbf{\bibinfo{volume}{78}},
  \bibinfo{pages}{060502} (\bibinfo{year}{2008}).

\bibitem[{\citenamefont{Takada and Chatterjee}(2003)}]{takada}
\bibinfo{author}{\bibfnamefont{Y.}~\bibnamefont{Takada}} \bibnamefont{and}
  \bibinfo{author}{\bibfnamefont{A.}~\bibnamefont{Chatterjee}},
  \bibinfo{journal}{Phys. Rev. B} \textbf{\bibinfo{volume}{67}},
  \bibinfo{pages}{081102} (\bibinfo{year}{2003}).

\bibitem[{\citenamefont{Clay and Hardikar}(2005)}]{clay}
\bibinfo{author}{\bibfnamefont{R.~T.} \bibnamefont{Clay}} \bibnamefont{and}
  \bibinfo{author}{\bibfnamefont{R.~P.} \bibnamefont{Hardikar}},
  \bibinfo{journal}{Phys. Rev. Lett.} \textbf{\bibinfo{volume}{95}},
  \bibinfo{pages}{096401} (\bibinfo{year}{2005}).

\bibitem[{\citenamefont{Hardikar and Clay}(2007)}]{hardikar}
\bibinfo{author}{\bibfnamefont{R.~P.} \bibnamefont{Hardikar}} \bibnamefont{and}
  \bibinfo{author}{\bibfnamefont{R.~T.} \bibnamefont{Clay}},
  \bibinfo{journal}{Phys. Rev. B} \textbf{\bibinfo{volume}{75}},
  \bibinfo{pages}{245103} (\bibinfo{year}{2007}).

\bibitem[{\citenamefont{{Fehske} et~al.}(2008)\citenamefont{{Fehske}, {Hager},
  and {Jeckelmann}}}]{fehske}
\bibinfo{author}{\bibfnamefont{H.}~\bibnamefont{{Fehske}}},
  \bibinfo{author}{\bibfnamefont{G.}~\bibnamefont{{Hager}}}, \bibnamefont{and}
  \bibinfo{author}{\bibfnamefont{E.}~\bibnamefont{{Jeckelmann}}},
  \bibinfo{journal}{EPL (Europhysics Letters)} \textbf{\bibinfo{volume}{84}},
  \bibinfo{pages}{57001} (\bibinfo{year}{2008}).

\bibitem[{\citenamefont{Craig et~al.}(2007)\citenamefont{Craig, Varney,
  Pickett, and Scalettar}}]{helen}
\bibinfo{author}{\bibfnamefont{H.~A.} \bibnamefont{Craig}},
  \bibinfo{author}{\bibfnamefont{C.~N.} \bibnamefont{Varney}},
  \bibinfo{author}{\bibfnamefont{W.~E.} \bibnamefont{Pickett}},
  \bibnamefont{and} \bibinfo{author}{\bibfnamefont{R.~T.}
  \bibnamefont{Scalettar}}, \bibinfo{journal}{Phys. Rev. B}
  \textbf{\bibinfo{volume}{76}}, \bibinfo{pages}{125103}
  (\bibinfo{year}{2007}).

\bibitem[{\citenamefont{{Koller} et~al.}(2004)\citenamefont{{Koller}, {Meyer},
  {Ono}, and {Hewson}}}]{koller}
\bibinfo{author}{\bibfnamefont{W.}~\bibnamefont{{Koller}}},
  \bibinfo{author}{\bibfnamefont{D.}~\bibnamefont{{Meyer}}},
  \bibinfo{author}{\bibfnamefont{Y.}~\bibnamefont{{Ono}}}, \bibnamefont{and}
  \bibinfo{author}{\bibfnamefont{A.~C.} \bibnamefont{{Hewson}}},
  \bibinfo{journal}{EPL (Europhysics Letters)} \textbf{\bibinfo{volume}{66}},
  \bibinfo{pages}{559} (\bibinfo{year}{2004}).

\bibitem[{\citenamefont{Koller et~al.}(2004)\citenamefont{Koller, Meyer, and
  Hewson}}]{koller2}
\bibinfo{author}{\bibfnamefont{W.}~\bibnamefont{Koller}},
  \bibinfo{author}{\bibfnamefont{D.}~\bibnamefont{Meyer}}, \bibnamefont{and}
  \bibinfo{author}{\bibfnamefont{A.~C.} \bibnamefont{Hewson}},
  \bibinfo{journal}{Phys. Rev. B} \textbf{\bibinfo{volume}{70}},
  \bibinfo{pages}{155103} (\bibinfo{year}{2004}).

\bibitem[{\citenamefont{Werner and Millis}(2007)}]{werner}
\bibinfo{author}{\bibfnamefont{P.}~\bibnamefont{Werner}} \bibnamefont{and}
  \bibinfo{author}{\bibfnamefont{A.~J.} \bibnamefont{Millis}},
  \bibinfo{journal}{Phys. Rev. Lett.} \textbf{\bibinfo{volume}{99}},
  \bibinfo{pages}{146404} (\bibinfo{year}{2007}).

\bibitem[{\citenamefont{{Bauer}}(2010)}]{bauer}
\bibinfo{author}{\bibfnamefont{J.}~\bibnamefont{{Bauer}}},
  \bibinfo{journal}{EPL (Europhysics Letters)} \textbf{\bibinfo{volume}{90}},
  \bibinfo{pages}{27002} (\bibinfo{year}{2010}).

\bibitem[{\citenamefont{{Bauer} and {Hewson}}(2010)}]{bauer2}
\bibinfo{author}{\bibfnamefont{J.}~\bibnamefont{{Bauer}}} \bibnamefont{and}
  \bibinfo{author}{\bibfnamefont{A.~C.} \bibnamefont{{Hewson}}},
  \bibinfo{journal}{\prb} \textbf{\bibinfo{volume}{81}}, \bibinfo{eid}{235113}
  (\bibinfo{year}{2010}).

\bibitem[{\citenamefont{Kumar and van~den Brink}(2008)}]{kumar}
\bibinfo{author}{\bibfnamefont{S.}~\bibnamefont{Kumar}} \bibnamefont{and}
  \bibinfo{author}{\bibfnamefont{J.}~\bibnamefont{van~den Brink}},
  \bibinfo{journal}{Phys. Rev. B} \textbf{\bibinfo{volume}{78}},
  \bibinfo{pages}{155123} (\bibinfo{year}{2008}).

\bibitem[{\citenamefont{Berger et~al.}(1995)\citenamefont{Berger,
  Val\'a\ifmmode~\check{s}\else \v{s}\fi{}ek, and von~der Linden}}]{berger}
\bibinfo{author}{\bibfnamefont{E.}~\bibnamefont{Berger}},
  \bibinfo{author}{\bibfnamefont{P.}~\bibnamefont{Val\'a\ifmmode~\check{s}\else
  \v{s}\fi{}ek}}, \bibnamefont{and}
  \bibinfo{author}{\bibfnamefont{W.}~\bibnamefont{von~der Linden}},
  \bibinfo{journal}{Phys. Rev. B} \textbf{\bibinfo{volume}{52}},
  \bibinfo{pages}{4806} (\bibinfo{year}{1995}).

\bibitem[{\citenamefont{Hotta and Takada}(1997)}]{hotta}
\bibinfo{author}{\bibfnamefont{T.}~\bibnamefont{Hotta}} \bibnamefont{and}
  \bibinfo{author}{\bibfnamefont{Y.}~\bibnamefont{Takada}},
  \bibinfo{journal}{Phys. Rev. B} \textbf{\bibinfo{volume}{56}},
  \bibinfo{pages}{13916} (\bibinfo{year}{1997}).

\bibitem[{\citenamefont{Johnston}(2010)}]{steve_thesis}
\bibinfo{author}{\bibfnamefont{S.}~\bibnamefont{Johnston}},
  \bibinfo{journal}{Ph.D thesis, University of Waterloo}
  (\bibinfo{year}{2010}).

\bibitem[{\citenamefont{Blankenbecler et~al.}(1981)\citenamefont{Blankenbecler,
  Scalapino, and Sugar}}]{BSS}
\bibinfo{author}{\bibfnamefont{R.}~\bibnamefont{Blankenbecler}},
  \bibinfo{author}{\bibfnamefont{D.~J.} \bibnamefont{Scalapino}},
  \bibnamefont{and} \bibinfo{author}{\bibfnamefont{R.~L.} \bibnamefont{Sugar}},
  \bibinfo{journal}{Phys. Rev. D} \textbf{\bibinfo{volume}{24}},
  \bibinfo{pages}{2278} (\bibinfo{year}{1981}).

\bibitem[{\citenamefont{White et~al.}(1989)\citenamefont{White, Scalapino,
  Sugar, Loh, Gubernatis, and Scalettar}}]{White}
\bibinfo{author}{\bibfnamefont{S.~R.} \bibnamefont{White}},
  \bibinfo{author}{\bibfnamefont{D.~J.} \bibnamefont{Scalapino}},
  \bibinfo{author}{\bibfnamefont{R.~L.} \bibnamefont{Sugar}},
  \bibinfo{author}{\bibfnamefont{E.~Y.} \bibnamefont{Loh}},
  \bibinfo{author}{\bibfnamefont{J.~E.} \bibnamefont{Gubernatis}},
  \bibnamefont{and} \bibinfo{author}{\bibfnamefont{R.~T.}
  \bibnamefont{Scalettar}}, \bibinfo{journal}{Phys. Rev. B}
  \textbf{\bibinfo{volume}{40}}, \bibinfo{pages}{506} (\bibinfo{year}{1989}).

\bibitem[{\citenamefont{Scalettar et~al.}(1989)\citenamefont{Scalettar,
  Bickers, and Scalapino}}]{scalettar_holstein}
\bibinfo{author}{\bibfnamefont{R.~T.} \bibnamefont{Scalettar}},
  \bibinfo{author}{\bibfnamefont{N.~E.} \bibnamefont{Bickers}},
  \bibnamefont{and} \bibinfo{author}{\bibfnamefont{D.~J.}
  \bibnamefont{Scalapino}}, \bibinfo{journal}{Phys. Rev. B}
  \textbf{\bibinfo{volume}{40}}, \bibinfo{pages}{197} (\bibinfo{year}{1989}).

\bibitem[{\citenamefont{Noack et~al.}(1991)\citenamefont{Noack, Scalapino, and
  Scalettar}}]{noack}
\bibinfo{author}{\bibfnamefont{R.~M.} \bibnamefont{Noack}},
  \bibinfo{author}{\bibfnamefont{D.~J.} \bibnamefont{Scalapino}},
  \bibnamefont{and} \bibinfo{author}{\bibfnamefont{R.~T.}
  \bibnamefont{Scalettar}}, \bibinfo{journal}{Phys. Rev. Lett.}
  \textbf{\bibinfo{volume}{66}}, \bibinfo{pages}{778} (\bibinfo{year}{1991}).

\bibitem[{\citenamefont{Veki\ifmmode~\acute{c}\else \'{c}\fi{}
  et~al.}(1992)\citenamefont{Veki\ifmmode~\acute{c}\else \'{c}\fi{}, Noack, and
  White}}]{vekic}
\bibinfo{author}{\bibfnamefont{M.}~\bibnamefont{Veki\ifmmode~\acute{c}\else
  \'{c}\fi{}}}, \bibinfo{author}{\bibfnamefont{R.~M.} \bibnamefont{Noack}},
  \bibnamefont{and} \bibinfo{author}{\bibfnamefont{S.~R.} \bibnamefont{White}},
  \bibinfo{journal}{Phys. Rev. B} \textbf{\bibinfo{volume}{46}},
  \bibinfo{pages}{271} (\bibinfo{year}{1992}).

\bibitem[{\citenamefont{Trivedi and Randeria}(1995)}]{trivedi}
\bibinfo{author}{\bibfnamefont{N.}~\bibnamefont{Trivedi}} \bibnamefont{and}
  \bibinfo{author}{\bibfnamefont{M.}~\bibnamefont{Randeria}},
  \bibinfo{journal}{Phys. Rev. Lett.} \textbf{\bibinfo{volume}{75}},
  \bibinfo{pages}{312} (\bibinfo{year}{1995}).

\end{thebibliography}

\end{document}